\documentclass[aps, prapplied, reprint]{revtex4-2}

\usepackage{amsmath}
\usepackage{graphicx}
\usepackage{dcolumn}
\usepackage{bm}

\begin{document}


\title{Mode mapping $Q>500\,000$ photonic crystal nanocavities using free carrier absorption}

\author{Karindra Perrier} 
\author{Jerom Baas}
\author{Sebastiaan Greveling}
\affiliation{Nanophotonics, Debye Institute for Nanomaterials Science, Utrecht University, Princetonplein 1, 3584 CC Utrecht, The Netherlands}
\author{ Ga\"{e}lle Lehoucq} 
\author{ Sylvain Combri\'{e}}
\author{ Alfredo de Rossi}
\affiliation{Thales Research and Technology, Route D\'{e}partementale 128, 91767 Palaiseau, France}
\author{ Sanli Faez }
\author{Allard P. Mosk}
\affiliation{Nanophotonics, Debye Institute for Nanomaterials Science, Utrecht University, Princetonplein 1, 3584 CC Utrecht, The Netherlands}


\begin{abstract}
We demonstrate a nonlinear photomodulation spectroscopy method to image the mode profile of a high-Q photonic crystal resonator (PhCR). This far-field imaging method is suitable for ultrahigh-Q cavities which we demonstrate on a Q = 619000 PhCR. 
We scan the PhCR surface with a 405~nm pump beam that modulates the refractive index by local thermal tuning, while probing the response of the resonance.
We enhance resolution by probing at high power, using the thermo-optical nonlinearity of the PhCR.
Spatial resolution of the thermo-optical effect is typically constrained by the broad thermal profile of the optical pump. 
Here we go beyond the thermal limit and show that we can approach the diffraction limit of the pump light. This is due to free carrier absorption that heats up the PhCR only when there is overlap between the optical pump spot and the optical mode profile.
This is supported with a thermo-optical model that reproduces the high-resolution mode mapping. Results reveal that the observed enhanced resolution is reached for surprisingly low carrier density. 

\end{abstract}

\maketitle



\section{\label{sec:1}Introduction}

Photonic crystals (PhC) have raised interest from fundamental research~\cite{lodahl2015interfacing, ghulinyan2015light}, due to their enhanced light-matter interaction, but maybe even more so from the field of optical signal processing and computing~\cite{baba2008slow, noda2003roadmap}. 
They are prime candidates for optical integrated-circuit applications~\cite{asano2018photonic, Caballero:22, moody2021roadmap}. 
One reason for this is the possibility to design a defect on the PhC platform, creating a PhC resonator (PhCR) that typically has small mode volume and high quality factor (Q)~\cite{wu2021nanoscale} which maximizes the optical nonlinearity crucial to active photonic devices.
Two-dimensional PhCs have been shown to be suitable for numerous applications such as fast optical switching \cite{husko2009ultrafast, nozaki2019femtofarad}, reconfigurable circuits~\cite{bruck2016all}, optical memories~\cite{Kuramochi2014} and optical parametric oscillation~\cite{marty2021photonic}.

Investigating the optical mode profile of the PhCR is vital in all of these functionalities. 
Calculations to predict resonator modes, like FDTD simulation, do not provide complete information as PhCR modes are very sensitive to nanometer-scale imperfections arising from the fabrication process~\cite{faggiani2016lower}. 
Therefore, high resolution imaging of PhCR modes is a necessary step in research and application development~\cite{rotenberg2014mapping}.

A benchmark method to characterize optical mode profiles is near-field scanning optical microscopy (NSOM), where a probe tip scans the surface of the PhC.
High spectral and deep subwavelength spatial resolution imaging have been demonstrated~\cite{rotenberg2014mapping, mujumdar2007near, knight1996characterizing}, as well as phase contrast imaging~\cite{schnell2010phase} and imaging of the electric and magnetic field~\cite{caselli2021near,kobus2014simultaneous}.
However, the requirement of nanometer-distances between the tip and the sample can be demanding, and makes this method unsuitable for devices that are covered by a top cladding. 
Additionally, perturbation by an NSOM probe tip will affect the Q-factor of the cavity~\cite{Cognee:19} making it challenging to image ultrahigh-Q cavities of order $10^4 - 10^6$ without destroying 
the confinement~\cite{arango2022cloaked, lalouat2008subwavelength}. Mujumdar et al. successfully show NSOM mode mappings of a Q=55000 cavity while noting that the imaging profoundly influences the spectral characteristics of the mode~\cite{mujumdar2007near}.
They measure a Q-factor degradation close to a factor of 2, which is in line with the experiment of Lalouat et al.~\cite{lalouat2008subwavelength}.
A perturbation of this order in the field of a Q~$\sim10^5$ cavity would mean the probe tip becomes the vastly dominating loss mechanism of the cavity, defeating the first-order perturbation principle.  

Far-field imaging with photomodulation spectroscopy (PMS) techniques introduce an excitation beam normal to the PhC surface that scans the PhC surface and perturbs the optical field by modulating the refractive index. 
The highest spatial resolution is achieved with electron or ion beam PMS scans, either with pulsed~\cite{brenny2016near, choi2013observation, kuttge2010ultrasmall} or continuous wave (cw)~\cite{mcgehee2017two} sources probing the resonance. 
It should be noted that exposure to ion and electron beams lowers the Q of the resonator, and in fact this technique is also being purposefully used to irreversibly alter or tune PhCs. 
Ultrafast PMS experiments were demonstrated by Bruck et al.~\cite{bruck2016all,BruckAndMuskens2015device}, where the mode profile is imaged (and tuned) with an ultraviolet beam. 
Their method is rooted in a shift of the refractive index due to the Drude-like dispersion of a dense free-carrier plasma.
This type of direct index perturbation needs a high concentration of carriers, 
and thus high-power optical excitation that can only be maintained on the picosecond scale due to risk of damage by photo-oxidation. The advantage then is the information in the time domain.
However, picosecond-pulse excitation is not suitable for high-Q resonators, where the photon lifetime exceeds the lifetime of the carriers. 

Thermal PMS is a method that is attractive due to its universality and ease of use. 
A pump beam is used to locally heat the PhC, and the resulting thermal change in refractive index leads to a shift in resonance frequency. 
Previously, our group has shown that with local thermal tuning the mode profile of a high-Q cavity can be recovered with a resolution that is limited by thermal diffusion~\cite{lian2016measurement,sokolov2015local, perrier2020thermo}. 
The effective spot size depends on the surrounding medium and it is generally much larger than the optical pump spot, so that even after deconvolution a moderate resolution remains~\cite{Lian:Profiles}.

In this paper we demonstrate enhanced spatial resolution for nonlinear PMS (NPMS), i.e., thermal photomodulation spectroscopy in the thermo-optical nonlinear regime.
This is due to free-carrier absorption (FCA) that only occurs when the pump and probe fields spatially overlap.
Using NPMS we measure the mode profile of ultrahigh-Q resonators without affecting the Q-factor. 

Our method reaches excellent sensitivity even with continuous wave excitation, because the high-Q resonator enhances both the probe field that heats the free carriers and the sensitivity with which the resulting thermal index shift is detected. 
In this way, the nonlocal thermo-optical effect amplifies a very small and local absorption term. 
This intrinsic amplification mechanism, leveraging the high Q of the resonator, makes it possible to operate at a very low pump power, thus avoiding Q-factor degradation or other damage to the sample.

We reproduce our results using a thermo-optical (TO) model that takes into account all TO sources, showing that a significant effect already occurs when only a few hundred free carriers are present in the resonator.


\section{\label{sec:2}Experiment and Theory}



\begin{figure}
\centering\includegraphics[width=0.47\textwidth,trim={6mm 0 1.2cm 0},clip]{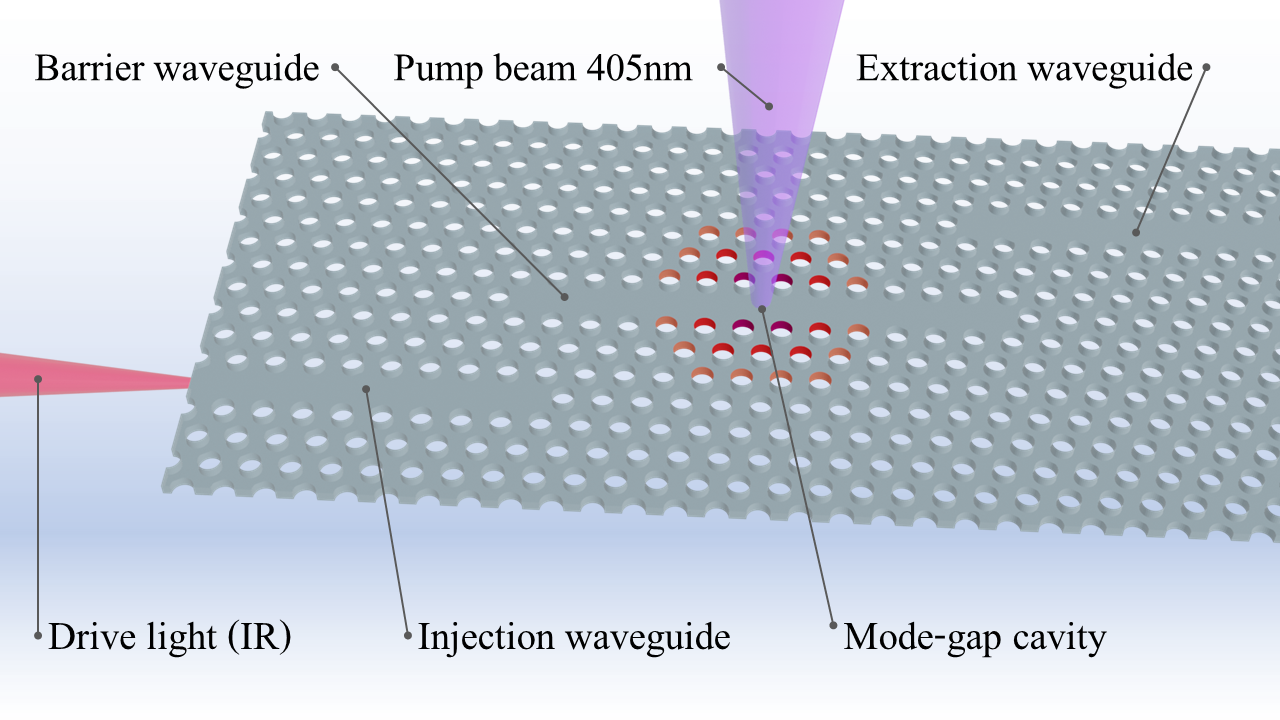}
\caption{Schematic of the GaInP PhC membrane, with lattice constant $a = 485$~nm, air hole radius of 136~nm, and membrane thickness of 180~nm. The mode-gap cavity consists of a broadened waveguide: the air holes lining the waveguide are shifted outwards from their lattice position (see colored air holes). We coupled NIR light in from the side into the injection waveguide, after which it evanescently couples to the barrier waveguide.}
\label{fig:sample}
\end{figure}

In Fig.~\ref{fig:sample} we show a schematic of the PhC, consisting of a 180~nm-thick GaInP slab with air holes in a triangular lattice and a lattice constant of $a = 485$~nm. 
The Ga$_{\mathrm{0.51}}$In$_{\mathrm{0.49}}$P slab was grown epitaxially on GaAs and has a bandgap of 1.9eV~\cite{schubert2018light} for TE polarized light. 
A nanocavity with a high Q and small mode volume was fabricated by waveguide width modulation, i.e., a mode-gap cavity~\cite{combrie2009high, kuramochi2006ultrahigh}.
TE polarized NIR probe light from a C-band tunable laser with a linewidth of 40MHz and accuracy of $\pm$1~pm is coupled into the injection waveguide with a polarization maintaining lensed fiber, and it then evanescently couples through to the barrier waveguide. 
A small fraction of light inside the cavity is scattered out of plane and collected with a 0.4 NA objective onto a cooled photodiode, this signal of out of plane scattered light is proportionate to the intracavity energy. 
The same objective focuses out-of-plane pump light onto the sample. 
The 405~nm pump light is steered by a fast scanning mirror, providing a localized heat source with which we scan the surface of the membrane. With the scanning pump beam we spatially probe the response of the cavity by measuring the intracavity energy via the out of plane scattered NIR light.


\begin{figure*}
\centering\includegraphics[width=0.83\textwidth,trim={4.1cm 2.2cm 4cm 1.8cm},clip]{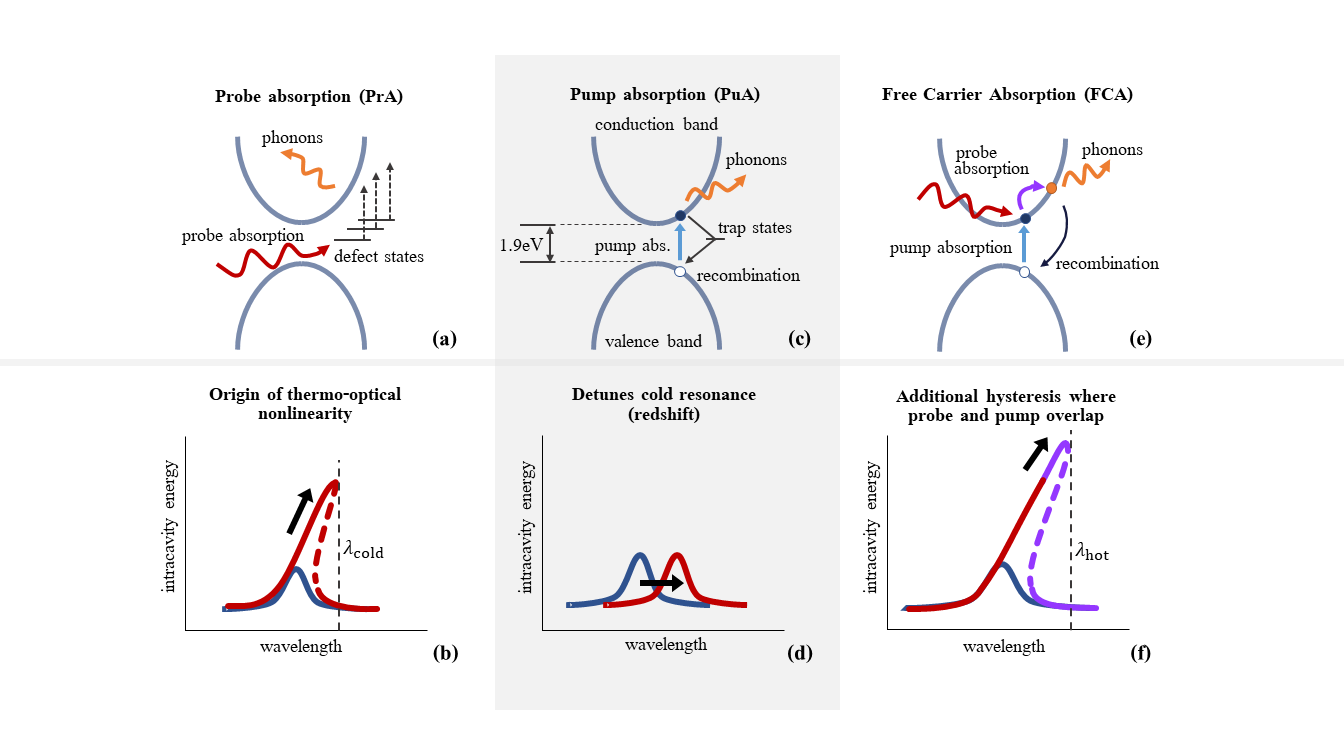}
\caption{ Schematic of the TO sources in the PhC. (a) NIR probe absorption by defect states causes (b) hysteresis behavior at high probe power (red lineshape). The blue lineshape illustrates the low-power, linear response. Dashed lines are instable branches. (c) Interband pump absorption of the 405~nm excitation laser generates a (d) shift of the resonance, i.e., a redshift of the complete lineshape, here depicted for the low-power lineshape. (e) Pump absorption creates free carriers that absorb NIR probe light, i.e., FCA (purple arrow), this causes an (f) increase in the hysteresis in the nonlinear lineshape (red and purple). 
The right hysteresis edge of the pumped nonlineare lineshape ($\lambda_{\mathrm{hot}}$ in (f)) is redshifted with respect to the hysteresis edge of the unpumped nonlinear lineshape ($\lambda_{\mathrm{cold}}$ in (b)).
}
\label{fig:sources}
\end{figure*}

In Fig.~\ref{fig:sources} we show a schematic of the three TO sources in the crystal. In Fig.~\ref{fig:sources}(a) we depict the processes leading to direct probe light absorption in the PhC. Since the bulk bandgap exceeds the probe photon energy by more than a factor two this takes place via surface states, impurities, free electrons in the semiconductor, or a combination of these processes. This probe absorption is the origin of the TO nonlinearity which causes the well-known hysteresis behavior shown in Fig.~\ref{fig:sources}(b)~\cite{perrier2020thermo, iadanza2020model}. Secondly, Fig.~\ref{fig:sources}(c,d) shows how direct heating by the pump causes a redshift of the cold cavity resonance $\lambda_0$ when the thermal profile overlaps with the optical mode profile of the resonance~\cite{lian2016measurement}. 
In Fig.~\ref{fig:sources}(e,f) a third source of heat presents itself when the free carriers generated by the pump absorb a probe photon, i.e., FCA. These hot carriers heat the lattice and increase the TO redshift, causing an increase in the hysteresis. 
The FCA process occurs only in the region where the pump spot and optical mode profile overlap, and is proportional to the product of pump and probe energy densities. 

We model the system by solving the coupled optical resonance condition and the partial differential equation for thermal diffusion for all three TO sources. 
We solve the differential equation numerically using the Sturm-Liouville theory, propagating the modes in an orthogonal base~\cite{hahn_heat_2012}.
See Appendix~\ref{AppxA:Theory} for a full treatment of the theoretical model.

A qualitatively similar model by Iadanza et al.~\cite{iadanza2020model} differs from ours as it approximates the temperature distribution with elliptical regions.
However, to interpret mode profile scans we need to specifically take into account the spatial degrees of freedom of the PhCR mode and use it to evaluate the temperature distribution in time and space on a high resolution grid.


\section{Results}

We perform NPMS measurements on a high-Q mode-gap cavity and image the optical mode profile. 
The procedure consists of a pump line scan: probing at consecutive pump positions in a horizontal or vertical line. At each pump position we take a probe wavelength sweep (in the direction of increasing $\lambda$) to obtain a cross section of the mode profile.
First we obtain both the theoretical and experimental response of the known mode profile of the fundamental mode-gap resonance with Q~=~619000 and validate the TO model.
Secondly, we demonstrate that the increased pump sensitivity due to FCA resolves the first higher-order (HO) mode of the mode-gap resonance with Q~=~340000 in the nonlinear regime.

\subsection{\label{sec:vertical_scan}Fundamental mode-gap resonance: vertical cross section }

\begin{figure*}
    \includegraphics[width=1\textwidth,trim={0 0 0 0},clip]{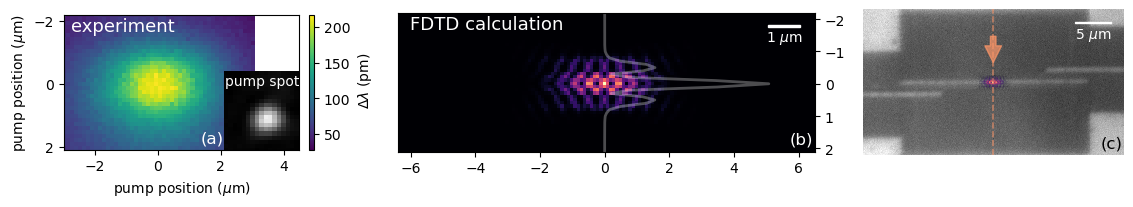}
    \caption{ Experimental and theoretical fundamental cavity mode. (a) Experimental 2D NPMS measurement of the mode, we plot $\Delta\lambda = \lambda_{\mathrm{hot}}-\lambda_{\mathrm{cold}}$ the pump induced redshift of the right hysteresis edge, see Figs.~\ref{fig:sources}(b,f). (b) FDTD calculated mode profile of the fundamental mode, with a vertical cross section through the center of the mode plotted in gray. This mode is used as input mode for the TO model. (c) Overlay of the input mode on top of a CCD image of the PhC waveguide. Orange dashed line and arrow indicate line scan and scan direction for the NPMS vertical cross section measurement shown in Fig.~\ref{fig:VerticalScan}. }
    \label{fig:InputModeVertical}
\end{figure*}

\begin{figure*}
    \centering
    \includegraphics[width=0.87\textwidth,trim={1mm 0 1mm 0},clip]{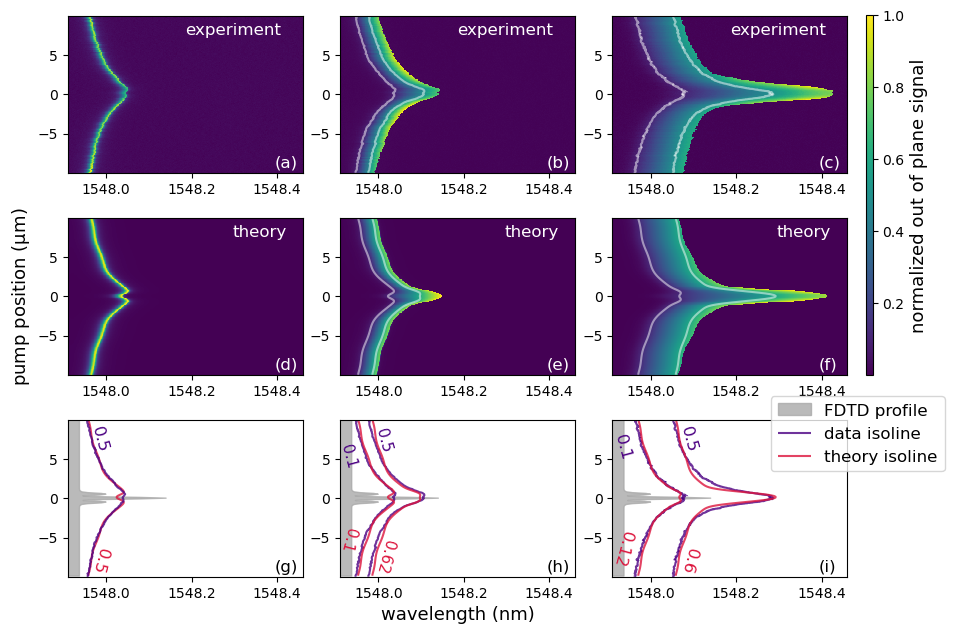}
    \caption{
    NPMS vertical cross section of the mode profile of a mode-gap cavity, experimentally measured via the out of plane scattered light at (a) the linear response regime at low input power, at (b) increasing input power and (c) high input power deep into the nonlinear response regime.
    The simulated NPMS mode profile in (d), (e) and (f) for increasing input power left to right, generated with the input mode of Fig.~\ref{fig:InputModeVertical}, matching the corresponding experimental results. See Appendix~\ref{AppxB:SimulationParameters} for fit details.
    (g,h,i) The isolines of the profiles, corresponding to the white isolines in (b,c,e,f).
    A cross section of the FDTD profile (corresponding to Fig.~\ref{fig:InputModeVertical}(c)) is indicated with a shaded area (gray).
    }
    \label{fig:VerticalScan}
\end{figure*}

Fig.~\ref{fig:InputModeVertical}(a) shows a two-dimensional NPMS measurement of the fundamental mode-gap cavity yielding an image of the mode profile. The NPMS measurement is performed with a 0.5~$\mu$m pump spot, the signal plotted is the pump induced redshift of the sharp right hysteresis edge of the cavity lineshape, see Fig.~\ref{fig:sources}(b,f).
The inset of the pump spot shows that the resolution of our measurement is close to the beam size of our spatial probe. 

Fig.~\ref{fig:InputModeVertical}(b) shows the optical mode profile of the 3D FDTD-simulated mode-gap resonance that we use to generate the theoretical NPMS obtained profile. The vertical line scan goes through the center of the mode-gap resonator as depicted in Fig.~\ref{fig:InputModeVertical}(c). 
While with our current pump spot size we do not experimentally resolve the standing wave fringes of the FDTD calculated mode, we do observe an elliptical mode profile.

In Fig.~\ref{fig:VerticalScan} we show the experimental and theoretical NPMS vertical line scan of the mode-gap resonance profile, performed with a 0.5~$\mu$m FWHM pump spot.
This was done for three increasing probe energies to investigate the response of the mode in both the linear and nonlinear regime, effectively increasing FCA. 
We observe a nearly perfect agreement between the experimentally measured and numerically calculated profiles.
A full list of the fit parameters and constants used in the model can be found in Appendix~\ref{AppxB:SimulationParameters}. All fits are found by manually adjusting four fit parameters, as an automated fit procedure failed to converge in a reasonable number of iterations, especially as each iteration requires a time-intensive calculation.

In Figs.~\ref{fig:VerticalScan}(a,d), at low probe power, pump absorption is the only significant TO effect. The resonance is redshifted by the pump where the mode profile and pump thermal profile overlap, i.e, when the pump is close to the cavity. 
A dip in the redshift when the pump hits the center of the waveguide is explained by considering the thermal conductance of the PhC. On the waveguide there are no air holes, causing a higher thermal conductance. Additionally, in the model we account for the observation that the pump is partially reflected by the waveguide (it shows up bright at the pump wavelength) diminishing the absorbed-pump power by 20\% when the pump beam is on the waveguide.  

In Figs.~\ref{fig:VerticalScan}(b,e), at increasing probe power, the TO effect causes an asymmetric lineshape (see Figs.~\ref{fig:sources}(b,f)) that maximally detunes the resonance when pump and probe heat up the mode cumulatively. 

In Figs.~\ref{fig:VerticalScan}(c,f) we probe deep into the non-linear regime. The high-energy-density areas of the mode profile become even more pronounced, as the TO detuning reaches further into the long wavelengths due to FCA. 
It is by these shapes that we scale the FCA effect to match the experiment and find an agreement with a carrier lifetime of $\tau_c\approx 0.4$~ns for all simulations throughout this article. The strength of the FCA source is related to the product of the carrier lifetime and the effective cross section, see Appendix~\ref{AppxA:Theory} for details. 

The carrier lifetime is the parameter with the largest a priori uncertainty since no measurements in these type of GaInP slabs exist to the best of our knowledge. Therefore we use it as a fit parameter. Our fit results should not be regarded as a measurement of the carrier lifetime, though it falls within the expected range for a GaInP PhC considering small volume to surface ratio~\cite{park_carrierlifetime_2014, holzman2005picosecond, kondo2013ultrafast,park_carrierlifetime_2014,thiagarajan1991picosecond}.

The lifetime translates to a carrier density of $\rho_c\approx 1\times 10^{16}$~cm$^{-3}$ or equivalently not more than $\sim 500$ carriers in the cavity at any time. 

In Figs~\ref{fig:VerticalScan}(g-i) low- and mid-level isolines of the experimental and theoretical profiles are shown.
Since the high intensity edge of the high-power profiles marks the transition to an unstable state (see lineshapes of Fig.~\ref{fig:sources}(b,f)) the edge is very sensitive to any effect that might disturb the balance at the edge of the stable energy branch, such as a nearby dark mode, variation in the incoupling efficiency, or disorder in the PhC.
Therefore, comparison of isolines that are situated in the stable region of the profile is more reliable. 
In Appendix~\ref{AppxC:SuppressedMode} we present another mode-gap resonance that has a suppressed high energy density peak of the mode profile, but nonetheless shows excellent agreement with the model in the stable regimes of the mode profile. 

Fig.~\ref{fig:VerticalScan} demonstrates that imaging of the PhCR mode improves at high probe power since the highly broadened FDTD profile in Fig.~\ref{fig:VerticalScan}(g) gains resolution in Fig.~\ref{fig:VerticalScan}(i).


\subsection{\label{sec:horizontal_scan}Higher order resonance: horizontal cross section}

\begin{figure*}
    \centering
    \includegraphics[width=0.66\textwidth,trim={1cm 0 0 0},clip]{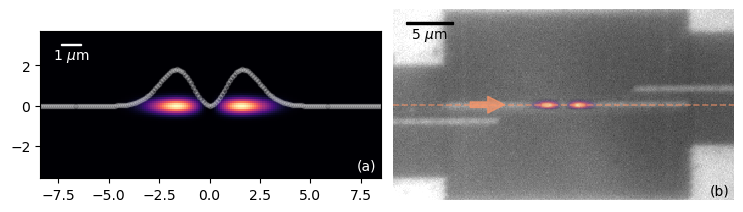}
    \caption{ Input mode for the theoretical NPMS horizontal line scan. (a) Render of the first higher-order Hermite-Gaussian function that is used as the input mode of the theoretical NPMS profile. A horizontal cross section of the input mode is plotted in gray. (b) Overlay of the input function on top of a CCD image of the PhC waveguide. The orange dashed line and arrow indicate the horizontal line scan and scan direction.  
    }
    \label{fig:InputModeHorizontal}
\end{figure*}

\begin{figure*}
    \centering
    \includegraphics[width=0.82\textwidth,trim={0 0 0 0},clip]{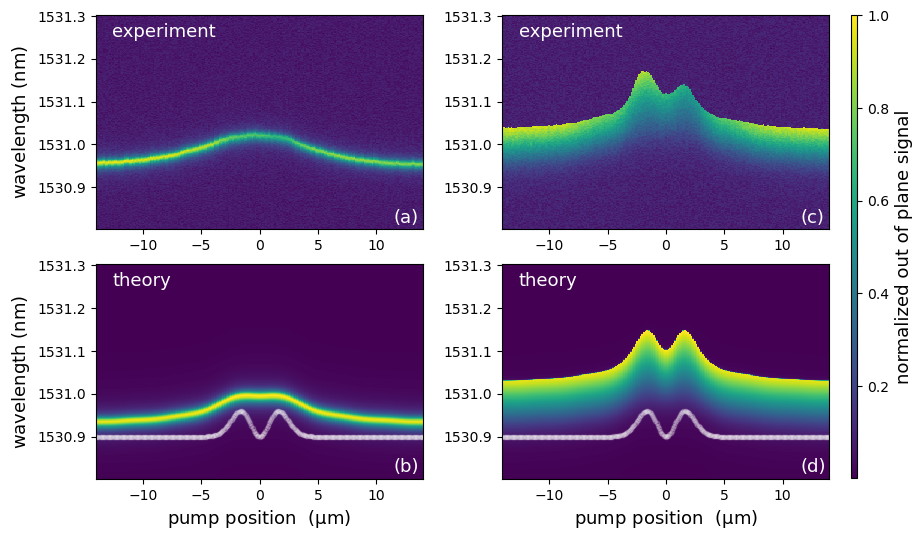}
    \caption{ 
    NPMS horizontal line scan of the first higher-order mode-gap resonance, yielding a cross section of the optical mode profile. (a) The profile is experimentally measured via out-of-plane scattered light at low probe power and (b) the theoretical profile is generated with the input mode from Fig.~\ref{fig:InputModeHorizontal} at low probe power. Additionally, (c) the experimentally measured profile at high probe power and (d) the theoretical profile at high probe power. For the theoretical profiles, a horizontal cross section of the input mode is plotted in white.
    }
    \label{fig:HorizontalScan}
\end{figure*}

Fig.~\ref{fig:InputModeHorizontal}(a) depicts the first HO mode of a Hermite-Gaussian function by which we approximate the first HO mode of the mode-gap resonance. 
We use this HO mode to generate the theoretical NPMS profile by performing a horizontal scan over the center of the waveguide as shown in Fig.~\ref{fig:InputModeHorizontal}(b).

Fig.~\ref{fig:HorizontalScan} shows the experimentally and theoretically obtained mode profile of the HO mode-gap resonance in both the low and high probe power regime. 
The low-probe power, experimental profile in Fig.~\ref{fig:HorizontalScan}(a) shows no details that would point to the measurement of a HO mode or any other mode that would have multiple peaks. 
In correspondence, the low-power theoretical profile in Fig.~\ref{fig:HorizontalScan}(b) shows the same single broad-peaked shape. 

However, at higher probe power the NPMS experimental mode profile in Fig.~\ref{fig:HorizontalScan}(c) reveals a centered double peak that highly resembles the double peak of the HO mode-gap profile modeled in Fig.~\ref{fig:HorizontalScan}(d). 
We see a very good agreement between experiment and theory is obtained by assuming this resonance is the first HO mode of the mode-gap resonance. We note that the characteristics of this HO mode can only be observed in the nonlinear regime at high probe power, where FCA is significant, proving again that our method enables high-resolution mode mapping of the resonance.

The distance between the peaks in Fig.~\ref{fig:HorizontalScan}(d) is fitted to match Fig.~\ref{fig:HorizontalScan}(c) by setting the waist of the HO Hermite-Gaussian mode to $w = 2.3\mu m$. 
We remark that the waist of the HO mode is of the same order as the fundamental mode, see Fig.~\ref{fig:InputModeVertical}.

We use the same fit method and parameters in the model as before in Fig.~\ref{fig:VerticalScan}, including the same carrier lifetime that scales the FCA source strength. See Appendix~\ref{AppxB:SimulationParameters} for a full list of constants and parameters.


\section{Discussion}

In our current experimental setup we do not reach the resolution limit of the imaging method. 
The most important limiting factor for the resolution is the pump focus, as illustrated in Fig.~\ref{fig:InputModeVertical}. Therefore we simulate with the realistically smallest pump spot of 213 nm (obtainable with an objective of NA = 0.95) and find a resolution limit of 240 nm using Sparrow's criterion, indeed approaching the diffraction limit of the 405~nm pump light itself. The resolution limit of 240 nm opens up prospects of measuring the fringes of the standing waves of the optical mode in Fig.~\ref{fig:InputModeVertical}(b). 
See Appendix~\ref{AppxD:Resolution} for details on the resolution limit simulation.

We note that, other than in NSOM measurements, we probe the field inside the PhC membrane as opposed to just above the PhC surface. With the absorption coefficient estimated at $9\times10^4$~\cite{adachi1992physical} the 180~nm thick slab will absorb 80$\%$ of the pump light with a slight bias towards the surface. Therefore our probing is determined by the cylindrical volume integral where absorbed pump light and mode profile overlap in the slab.


\section{Conclusion}

We provide a simple, non-invasive approach to image optical mode profiles in the far-field, with its own unique trade-off between spatial and spectral resolution, reproducibility, and applicability.
We demonstrated that NPMS can be used to obtain a high-resolution mode mapping of ultrahigh-Q PhC resonances by driving the optical mode into the TO nonlinear regime. 
We push the boundaries of far-field imaging~\cite{de2010optical} and resolve the fringes in the profile of a HO mode. 

The TO model we present shows that FCA, invoked by high-power probing of the mode, is the origin of the augmented resolution, as pump and probe light have an accumulative effect on the detuning of the resonance. In this way, we beat the convolution effects of the broad thermal profile of the pump spot. This is surprising at the low carrier density of $\approx1\times10^{16}$~cm$^{-3}$, as typical carrier index modulation experiments require densities on the order of $10^{18-20}$~cm$^{-3}$~\cite{fan1967effects, BruckAndMuskens2015device} and thus pulsed light. 

NPMS offers a convenient cw low pump-power method for for instance silicon photonic circuits and devices designed to combine high-Q dielectric resonators with plasmonic systems~\cite{chen2018hybrid}.
Additionally, it can provide insight into the field inside the dielectric material, measuring the mode profile within the PhC slab. 
Our method can easily be expanded to time-resolved measurements. 
Moreover, opportunities to dynamically tune and configure modes --- possibly using opposite-sign index modulation of the different nonlinear effects in the material --- are worthy of exploring.


\begin{acknowledgments}
The authors would like to thank Aron Opheij, Dante Killian, Cees de Kok, Paul Jurrius and Aquiles Carattino for their technical support. 
This research is supported by the Dutch Research Council (NWO) (Vici 68047618).
\end{acknowledgments}





\bibliographystyle{apsrev4-1}
\bibliography{FCAModeMapping_Bib.bib}


\appendix{}


\section{Theory}\label{AppxA:Theory}

In this section we formulate a thermo-optical (TO) model 
by solving the optical resonance condition coupled to the thermal
diffusion equations for all three TO sources.
This means we will find an expression for the TO sources and work towards the full heat equation of the system, taking into account the optical mode in the waveguide, the thermal properties of the PhC and the optical pump.


\subsubsection{Carrier absorption}

First we will find an expression for the heat produced by free carrier absorption (FCA) as depicted in Fig.~\ref{fig:sources}(e), i.e., free carriers produced by the pump light absorb probe light. 
To find the thermal dissipation from the FCA source we multiply the optical absorption coefficient with the energy density of the probe field. This gives us the energy transfer from probe light to free carriers, i.e., the FCA heating term,

\begin{align}
  S_{\mathrm{FCA}}  &= \alpha_{\mathrm{FCA}}(x,y,t) \nonumber \\
                    &\hspace{0.8cm} \times \left( \frac{c}{n} E_{\mathrm{pr}}(t) U_{\mathrm{pr}}(x,y) I_{\mathrm{pr}}(x,y,t) \right),
\end{align}

\noindent
where 
$\alpha_{\mathrm{FCA}}(x,y)$ is the probe absorption coefficient, 
$c$ is the speed of light, $n$ the refractive index,
$E_{\mathrm{pr}}(t)$ is the input probe light energy, $U_{\mathrm{pr}}(x,y)$ the optical mode profile of the probe field, and $I_{\mathrm{pr}}(x,y,t)$ is the spectral lineshape of the resonance.

The probe energy is determined by the input probe power $P_{\mathrm{pr}}$ and a coupling parameter $\tau_{\mathrm{pr}}$ accounting for the incoupling efficiency of the probe light to the cavity, which means $E_{\mathrm{pr}}(t) = \tau_{\mathrm{pr}} P_{\mathrm{pr}}(t)$.
The absorption coefficient is related to the effective FCA cross section $\sigma_{\mathrm{FCA}}$ and the carrier density $\rho_c(x,y)$ by $\alpha_{\mathrm{FCA}}(x,y,t) = \sigma_{\mathrm{FCA}} \rho_c(x,y,t)$. 
Where the effective absorption cross section is the summed cross sections of all carrier types, where for each type $ \sigma_i = e^3 \lambda_{\mathrm{pr}}^2/4 \pi^2 c^3 (m_i^*)^2 \mu_i n \varepsilon_0 $ is the absorption cross section of a single carrier \cite{li2006excess}. Here $e$ is the electron charge, $\lambda_{\mathrm{pr}}$ the probe wavelength, $m^*$ the effective carrier mass, $\mu$ is the carrier mobility and $\epsilon_0$ the vacuum permittivity.

The carriers are created by the pump light. We assume the carriers stay localized on the optical timescale \cite{park_carrierlifetime_2014}, which means the carrier density is directly proportional to the local production rate of carriers and their lifetime, such that 
\begin{equation}
    \rho_{c}(x,y,t) = \frac{\tau_c}{d \hbar \omega_{\mathrm{pu}}} P_{\mathrm{pu}}(t)U_{\mathrm{pu}}(x,y),    
    \label{eq:carrierdensity}
\end{equation}
with $\tau_c$ the average lifetime of a carrier, $d$ the thickness of the membrane, $\omega_{\mathrm{pu}}$ the pump light frequency, $P_{\mathrm{pu}}(t)$ the pump power, and $U_{\mathrm{pu}}(x,y)$ the optical mode profile of the pump.
This finally leads to the expression

\begin{align}
  S_{\mathrm{FCA}} 
                &= \sigma_{\mathrm{FCA}} \left( \frac{\tau_c }{d \hbar \omega_{\mathrm{pu}}} P_{\mathrm{pu}}(t)U_{\mathrm{pu}}(x,y) \right) \nonumber\\
                & \hspace{1.1cm} \times \left( \frac{c\tau_{\mathrm{pr}}}{n} P_{\mathrm{pr}}(t) U_{\mathrm{pr}}(x,y)I_{\mathrm{pr}}(x,y,t) \right).
\label{eq:FCAsource}
\end{align}


\subsubsection{Heat Equation}

The heat equation, taking into account all TO sources (shown in Fig.~\ref{fig:sources}), is a partial differential equation which we solve using a stable spectral method on a high-resolution space-time grid. We combine the in-plane cooling through dissipation in the PhC membrane, out-of-plane cooling via the surrounding gas layer and substrate, and all heating terms, including the FCA term of Eq.~\eqref{eq:FCAsource}, into the heat equation

\begin{eqnarray}
  C_{\mathrm{2D}} &&  \frac{\partial T(x,y,t)}{\partial t} =\\
  &&\hspace{0.5cm} - K_{\mathrm{2D}} \nabla^2 T(x,y,t)\nonumber\\
  &&\hspace{0.5cm} - K_{\mathrm{gas}} (T(x,y,t)-T_0) \nonumber\\
  &&\hspace{0.5cm} + \alpha_{\mathrm{PrA}} P_{\mathrm{pr}}(t)U_{\mathrm{pr}}(x,y) \nonumber\\
  &&\hspace{0.5cm} + \beta_{\mathrm{PuA}} P_{\mathrm{pu}}(t)U_{\mathrm{pu}}(x,y) \nonumber\\
  &&\hspace{0.5cm} + \sigma_\mathrm{FCA} 
  \frac{\tau_{\mathrm{c}}}{d \hbar \omega_{\mathrm{pu}}}P_{\mathrm{pu}}(t)U_{\mathrm{pu}}(x,y) \nonumber \\
  && \hspace{2.0cm}\times \frac{c\tau_{\mathrm{pr}}}{n}P_{\mathrm{pr}}(t)U_{\mathrm{pr}}(x,y)I_{\mathrm{pr}}(x,y,t),\nonumber
  \label{SMeq:heatequation}
\end{eqnarray}

\noindent 
where $C_{\mathrm{2D}}$ is the 2D specific heat of the PhC membrane, $K_{\mathrm{2D}}$ the thermal conductivity of the PhC membrane, and $K_{\mathrm{gas}}$ the thermal conductance of the gas layer between the air-suspended membrane and substrate. At the boundary we have $T = T_0$ the temperature at the edge of the photonic crystal where it meets the substrate. Out-of-plane cooling via the other side of the PhC membrane into free space is neglected, as this contributes less then one percent of the heat loss. 
For the heat produced via probe absorption we take into account the absorption fraction $\alpha_{\mathrm{PrA}}$, i.e., the probability a cavity photon is absorbed rather than scattered or leaked out of the cavity.
Direct pump heating is related to $\beta_{\mathrm{PuA}}$, the absorption fraction of the pump light.

The resonance condition is hidden in the optical energy density of the probe field that is related to the spectral lineshape of the resonance

\begin{eqnarray}
    I_{\mathrm{pr}}(x,y,t) = \frac{\Gamma^2}{\Gamma^2+(\Delta-\delta_{\mathrm{th}}(x,y,t) )^2},
    \label{eq:resonancecondition}
\end{eqnarray}
\noindent where $\Gamma$ is the resonance linewidth, $\Delta \equiv \lambda_{\mathrm{pr}} - \lambda_0$ is the detuning, and $\delta_{\mathrm{th}}(t)$ is the thermal resonance shift that causes a dynamic detuning expressed by
\begin{equation}
    \delta_{\mathrm{th}}(x,y,t)= \eta \int U_{\mathrm{pr}}(x,y)  T(x,y,t) dx dy.
    \label{eq:delta}
\end{equation}

\noindent Here, $\eta$ is the thermo-optical coefficient of the semiconductor material.
Eq.~\eqref{eq:resonancecondition} reflects the fact that the resonance has a dynamic Lorentzian lineshape that detunes in accordance with the temperature.
For our experimental conditions the resonance frequency detunes linearly with temperature~\cite{Sokolov:Thermo-opticalCoeff}. 
The optical mode profile of the resonance, as depicted in Fig.~\ref{fig:InputModeVertical}(b) and Fig.~\ref{fig:InputModeHorizontal}(a), is inserted into $U_{\mathrm{pr}}(x,y)$. 
We then solve Eq.~\eqref{SMeq:heatequation} numerically using Sturm-Liouville theory, propagating the modes in an orthogonal base~\cite{hahn_heat_2012}. 


 
\section{Simulation parameters} \label{AppxB:SimulationParameters}

\begin{table*}
\centering
\begin{tabular}{l|l|l|l}
\multicolumn{4}{c}{Constants and parameters used in the model} \\ \hline \hline
  & symbol & value & source \\
 \hline
 refractive index               & $n$                             & 3.06      & \cite{goldberg1999gallium}\\
 thermal conductivity of GaInP  & $\kappa_{\mathrm{GaInP}}$     & 4.9 W/m.K     & \cite{Adachi:ThermalConductivity}\\
 thermal conductivity of N$_2$  & $\kappa_{\mathrm{N_2}}$ & 0.024 W/m.K   & \cite{ThermalConductivityNitrogen} \\ 
 specific heat of GaInP         & $C_{\mathrm{sp}}$             & 310 J/K.kg    & \cite{Piesbergen:SpecificHeatInP}  \\
 density of GaInP               &$\rho$                         & 4810 kg/m$^3$ &  \cite{density}\\
 thickness of the PhC membrane  &$h$                            & 180 nm        & \\
 filling fraction               &$\phi(x,y)$                    & 0.714 (PhC membrane) \\ 
                                &                               &1.0 (waveguide, bulk)        & \\
 2D specific heat               &$C_{\mathrm{2D}}(\phi(x,y))$   & $h\phi(x,y)\rho C_{\mathrm{sp}}$   &  \\
 thermo-optical coefficient     &$\eta = dT/dn$                        & $-2\times 10^{-4}$ K$^{-1}$    & \cite{Sokolov:Thermo-opticalCoeff}\\
 on-chip upper limit of pump power &$P_{\mathrm{lim}}$          & 1.56 $\mu$W    & experimentally measured \\
 on-chip pump power             &$P_{\mathrm{pu}}$            & 0.35 $\times P_{\mathrm{lim}}$ & fit parameter\\
 absorption fraction of probe   &$\alpha_{\mathrm{PrA}}$        & 0.05          & estimated value\\
 absorption fraction of pump    &  $\beta_{\mathrm{PuA}}$       & 1.0 (PhC membrane) \\
                                &                               & 0.2 (waveguide) & fit parameter  \\
 electron effective mass        & $m_{e}^*$     &   0.088$m_0$      & \cite{goldberg1999gallium}\\
 light hole effective mass      & $m_{lh}^*$    &   0.12$m_0$       & \cite{goldberg1999gallium}\\
 heavy hole effective mass      & $m_{hh}^*$    &   0.7$m_0$        & \cite{goldberg1999gallium}\\
 electron mobility              & $\mu_e$       &   1000 cm$^2$/Vs      & \cite{goldberg1999gallium}\\
 hole mobility                  & $\mu_h$       &   40 cm$^2$/Vs        & \cite{goldberg1999gallium}\\
 effective FCA cross section    & $\sigma_{\mathrm{FCA}}$ & $\sigma_{e} + 1/2(\sigma_{lh} + \sigma_{hh})$ \\
 
 Vertical line scan, Fig.~\ref{fig:VerticalScan}: &&\\
 \hspace{0.5cm} resonance wavelength    &$\lambda_0$        & 1547.960 nm   & experimentally measured \\
 \hspace{0.5cm} resonance linewidth     &$\Gamma$           & 5 pm          & experimentally measured \\
 \hspace{0.5cm} FWHM of the pump in (x,y)-direction     &   & (534, 587)~nm & experimentally measured \\
 \hspace{0.5cm} carrier lifetime    & $\tau_c$      & 0.4 ns  & fit parameter \\
 \hspace{0.5cm} probe light energy &$E_{\mathrm{pr}}=\tau_{\mathrm{pr}}P_{\mathrm{pr}}$ & 0.2 fJ & fit parameter Fig.~\ref{fig:VerticalScan}(d)\\
 \hspace{0.5cm}      &          & 1.6 fJ & fit parameter Fig.~\ref{fig:VerticalScan}(e)\\
 \hspace{0.5cm}      &          & 3.6 fJ & fit parameter Fig.~\ref{fig:VerticalScan}(f)\\
 Horizontal line scan, Fig.~\ref{fig:HorizontalScan}: &&\\
 \hspace{0.5cm} resonance wavelength                & $\lambda_0$           & 1530.952 nm   & experimentally measured \\
 \hspace{0.5cm} resonance linewidth                 & $\Gamma$              & 9 pm          & experimentally measured \\
 \hspace{0.5cm} FWHM of the pump in (x,y)-direction &                       & (615, 548)~nm  & experimentally measured\\
 \hspace{0.5cm} carrier lifetime    & $\tau_c$      & 0.4 ns  & fit parameter \\
 \hspace{0.5cm} probe light energy &$E_{\mathrm{pr}}=\tau_{\mathrm{pr}}P_{\mathrm{pr}}$ 
                                                                            & 0.1 fJ   & fit parameter Fig.~\ref{fig:HorizontalScan}(b)\\
 \hspace{0.5cm}         &                                                   & 2.1 fJ    & fit parameter Fig.~\ref{fig:HorizontalScan}(d)\\ 
 Vertical line scan, Fig.~\ref{fig:VerticalScan_Appdx}: &&\\
 \hspace{0.5cm} resonance wavelength                & $\lambda_0$   & 1536.106 nm   & experimentally measured \\
 \hspace{0.5cm} resonance linewidth                 & $\gamma$      & 4 pm          & experimentally measured \\
 \hspace{0.5cm} FWHM of the pump in (x,y)-direction &                & (615, 548)~nm  & experimentally measured\\
 \hspace{0.5cm} carrier lifetime    & $\tau_c$      & 0.4 ns  & fit parameter \\
 \hspace{0.5cm} probe light energy &$E_{\mathrm{pr}}=\tau_{\mathrm{pr}}P_{\mathrm{pr}}$ 
                                                                            & 0.3 fJ   & fit parameter Fig.~\ref{fig:VerticalScan_Appdx}(d)\\
 \hspace{0.5cm}         &                                                   & 1.7 fJ    & fit parameter Fig.~\ref{fig:VerticalScan_Appdx}(e)\\ 
 \hspace{0.5cm}         &                                                   & 6.6 fJ    & fit parameter Fig.~\ref{fig:VerticalScan_Appdx}(f)\\ 
\end{tabular}
\caption{ Constants and parameters used to generate the theoretical mode profiles of Figs.~\ref{fig:VerticalScan}(d,e,f), \ref{fig:HorizontalScan}(b,d) and Fig.~\ref{fig:VerticalScan_Appdx}(d,e,f). }
\label{tab:parameters}
\end{table*}

\begin{figure*}
    \centering
    \includegraphics[width=0.87\textwidth,trim={0 0 0 0},clip]{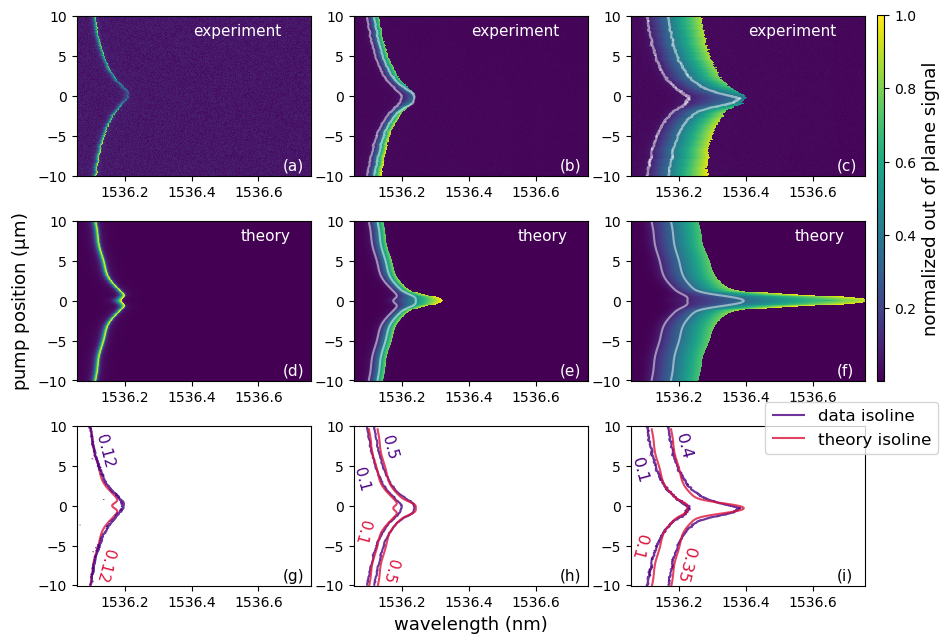}
    \caption{ 
        NPMS vertical cross section of the mode-gap cavity profile, experimentally measured at (a) low input power, (b) increasing input power and (c) high input power.
        The simulated NPMS mode profile in (d), (e) and (f) for increasing input power, generated with the input mode of Fig.~\ref{fig:InputModeVertical}.
        (g,h,i) The isolines of the profiles.
    }
    \label{fig:VerticalScan_Appdx}
\end{figure*}

Table~\ref{tab:parameters} shows the parameters and constants used in the model to generate Figs.~\ref{fig:VerticalScan}(d,e,f), \ref{fig:HorizontalScan}(b,d), and \ref{fig:VerticalScan_Appdx}(d,e,f). 

The on-chip pump power cannot be directly measured. We measure the upper limit of the on-chip pump power, scale to fit the redshift of the cold resonance observed in the experiment, and then fixed for all simulations. The shift of the cold resonance (in the linear response regime of the cavity) is directly related to the pump power and has no crosstalk with the other fit parameters. 

The probe energy $E_{\mathrm{pr}}$ determines the detuning at which the resonance drops out of stability, i.e., the high intensity edge. Therefore, the probe energy is a fit parameter to match the experimental hysteresis edge at pump position $\pm 10\mu m$ where the pump has negligible effect on the optical mode. The spectrum of the input probe energy is not entirely flat, therefore we normalize it using the back reflected light measured through the injection waveguide. 

The diminished pump absorption on the waveguide (caused by partial reflection on the waveguide) is determined in the vertical scan by fitting the fringe on top of the waveguide, see Fig.~\ref{fig:VerticalScan}(a,d). 

The fourth and last fit parameter is related to the strength of the FCA term, i.e., the amount of heat produced by FCA. Looking at Eq.~\eqref{eq:FCAsource} the FCA source is proportionate to the product $\sigma_{\mathrm{FCA}}\tau_c$, the effective FCA cross section and the carrier lifetime. 
We use the effective carrier mass and mobility 
to calculate the effective cross section and use the carrier lifetime as fit parameter to match the resolution of the experimental mode profile. A small carrier lifetime leads to a broadened profile with less sharp peaks and vice versa.

Using the values from Table~\ref{tab:parameters} the fit yields a carrier lifetime of $\tau_c \approx 0.4$~ns. This is an approximated value, as the effective cross section has large uncertainty due to the fact that the population of heavy and light carrier holes is unknown and we work with a 50/50 distribution. 
If only light holes are part of the FCA process the carrier lifetime would average out to $\tau_c\approx 0.2$~ns.
Additionally, the effective carrier mass and carrier mobility cannot directly be measured in our experimental setup, thus we use literature values with limited precision. With Eq.~\eqref{eq:carrierdensity} we find the carrier density created by the cw pump spot to be 
$\rho_c \simeq \tau_c P_{\mathrm{pu}}/\hbar\omega_{\mathrm{pu}}d\pi r^2 $, where $r$ is half the FWHM of the pump spot. This leads to the carrier density of $\rho_c \approx 1\times10^{16}$~cm$^{-3}$, or equivalently, not more than $\sim$500 carriers in the PhCR at any time.


\section{Suppressed FCA response}\label{AppxC:SuppressedMode}

At high optical probe energy the high energy branch of a bistable state can be susceptible to many disturbances that destabilize the branch. 
This is demonstrated in Fig.~\ref{fig:VerticalScan_Appdx} where we show a resonance that is suppressed in its nonlinear behavior as we do not see states with large detuning when we have overlap between pump and mode, due to an instable high energy branch. Nevertheless, the isolines in the stable part of the high energy branch (at low and medium intracavity energy) still match the theory.

\begin{figure*}
    \centering
    \includegraphics[width=0.86\textwidth,trim={0 0 0 0},clip]{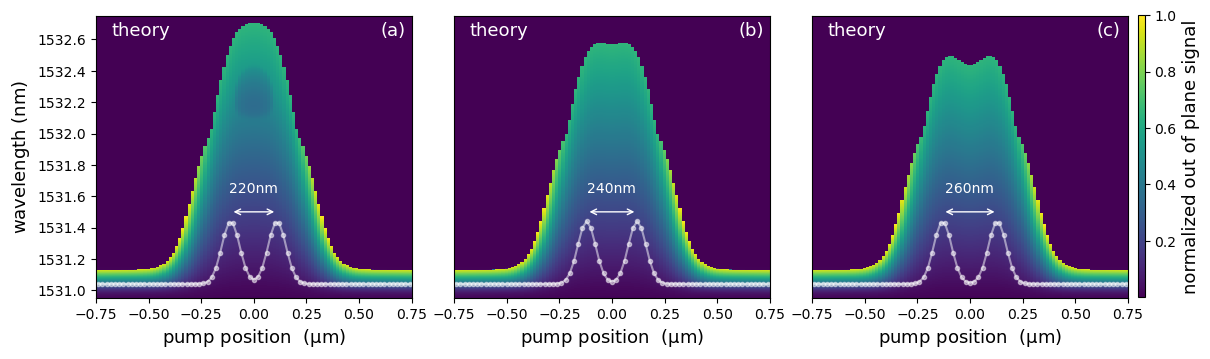}
    \caption{ 
        Simulation of the resolution limit using Sparrow's criterion. Increasing distance between Gaussian peaks: (a) 220, (b) 240, (c) 260 nm. Input modes are plotted in white. At (b) 240~nm the two peaks can still be resolved. 
    }
    \label{fig:Resolution_Appdx}
\end{figure*}

\section{Resolution limit simulation}\label{AppxD:Resolution}

The most important limiting factor for the resolution is the size of the pump beam, determined by the pump light wavelength and NA of the focusing objective via the diffraction limit $\lambda/2\mathrm{NA}$. The current setup has NA = 0.4 and $\lambda_{\mathrm{pump}} = 405$~nm. Objectives that work for both the NIR and pump wavelength go up to NA = 0.95. 
In principle the imaging method could be performed with a shorter wavelength pump light to obtain higher resolution. The limited transmission spectrum for a corresponding objective might mean the NIR light must be detected via the PhC waveguide transmission signal instead. 

To calculate the resolution limit achievable with our current 405~nm pump light we use a very tight but realistic pump focus of 213~nm that can be obtained with an objective of NA = 0.95. 
We then generate the mode profile for two sharp, adjacent Gaussian modes with FWHM = 100 nm and varying distance on the x-axis between them, using our TO model. 

Fig.~\ref{fig:Resolution_Appdx} shows the result of the simulation demonstrating that the resolution limit lies at 240~nm followings Sparrow's criterion. 
All constants and parameters are identical to the horizontal scan of Fig.~\ref{fig:HorizontalScan} except for an adjusted pump FWHM of 213 nm in both the x and y direction (see Table~\ref{tab:parameters}). 

The simulation does not include carrier diffusion, which might become significant at the nanometer scale. This is highly dependent on the unknown carrier lifetime. 
We estimate the diffusion coefficient at $D_e=26~cm^2/s$ and $D_h=1~cm^2/s$ for the electrons and holes respectively, using the Hall mobility and Einstein relation. 
The diffusion length is $\sqrt{<r^2>_i} = \sqrt{4D_i\tau_c}$ for each carrier type in the 2D slab. For a carrier lifetime of $\tau_c = \{ 10, 100, 400 \}$~ps we have diffusion lengths of $\{321,1016,2033\}$~nm and $\{64,203,407\}$~nm for the electrons and holes respectively.








\end{document}